\documentclass[12pt]{article} 

\usepackage{amsmath,amssymb,epsfig}

\def\etal{\emph{et al} }

\newcommand{\eps}{\varepsilon}

\def\be{\begin{equation}}
\def\ee{\end{equation}}

\def\a{\alpha}
\def\b{\beta}
\def\G{\Gamma}
\def\Gv{\Gamma v}
\def\ehat{\hat{\mathbf{e}}}

\def\la{\langle}
\def\ra{\rangle}     

\def\udot{\dot{u}}

\begin{document}

\begin{center}
{\large\bf

Kinematic and Weyl singularities}

\

W C Lim, A A Coley, S Hervik

Department of Mathematics \& Statistics, Dalhousie University,
Halifax, Nova Scotia, Canada B3H 3J5

Email: wclim@mathstat.dal.ca, aac@mathstat.dal.ca, herviks@mathstat.dal.ca

\

\today
\end{center}

\begin{abstract}

We study the properties of a future singularity encountered by a
perfect fluid observer in tilting spatially homogeneous Bianchi cosmologies.
We derive the boost formulae for the Weyl tensor to establish that, for 
two observers that are asymptotically null with respect to each other, 
their respective Weyl parameters generally both tend to zero, constant, or 
infinity together.
We examine three classes of typical examples and one exceptional class. 
Given the behaviour of the Weyl parameter, we can predict that the
singularity encountered is a Weyl singularity or a kinematic singularity.
The analysis suggests that the kinematic variables are also useful
in indicating a singularity in these models.

\
\end{abstract}

\section{Introduction}

A cosmological model ($\mathcal{M},\mathbf{g},\mathbf{u}$) is defined by specifying the 
spacetime geometry, determined by a Lorentzian metric $\mathbf{g}$ defined on the 
manifold $\mathcal{M}$, and a family of fundamental observers, whose congruence%
\footnote{A congruence is a family of curves such that through each point there passes 
 precisely one curve in this family.}
of world-lines is represented by the 4-velocity field $\mathbf{u}$. The covariant 
derivative $u_{a;b}$ of the 4-velocity field is decomposed into \emph{kinematic} 
variables according to
\be
	u_{a;b} = \sigma_{ab} + \omega_{ab} + H (g_{ab} + u_a u_b) - \dot{u}_a u_b,
\ee
where $\sigma_{ab}$ is the rate of shear tensor, $\omega_{ab}$ is the vorticity tensor, 
$H$ is the Hubble scalar, and $\dot{u}_a$ is the acceleration vector.
Sometimes there is another preferred family of fundamental observers, 
$\hat{\mathbf{u}}$, whose covariant derivative defines a corresponding set of kinematic 
variables $(\hat{\sigma}_{ab}, \hat{\omega}_{ab}, \hat{H}, \hat{\dot{u}}_a)$, which are 
different from the first set.

The study of spatially homogeneous (SH) Bianchi cosmologies forms 
part of the effort to understand the underlying dynamics of anisotropic 
and inhomogeneous universes. 
In recent papers \cite{CHL06,CHL06b}
we examined the future asymptotic dynamics as experienced by the
perfect fluid observer in Bianchi cosmologies 
with a tilted perfect fluid with linear equation of state $p=(\gamma-1)\mu$, 
focussing on physical properties of the models as seen by the fluid observer.
We found that given a Bianchi type, for a large enough value of the 
equation
of state parameter, $\gamma$, the perfect fluid observer can encounter a 
singularity, in spite
of the fact that the SH observers, moving geodesically and orthogonally to
the spatially homogeneous hypersurfaces, do not
encounter any singularity into the future \cite{Rendall}.
Although there is no universally accepted definition of a singularity~\cite[Section 
9.1]{Wald}, we shall make use of the congruence in defining a singularity within the context of a 
cosmological model: a singularity is encountered by a congruence if 
the proper time is finite and inextendible. This definition is consistent with the general definition 
adopted in the early work on Bianchi cosmologies by Ellis \& King, which states that a singularity is 
recognized by the existence of an inextendible curve which has a finite length as measured by a 
generalized affine parameter \cite[page 121]{EK74}.
\footnote{Many also use the stricter 
definition of a singularity as geodesic incompleteness. However, in our case this is not satisfactory 
since the fluid does not follow geodesics; thus we will not require the congruence to be geodesic.}
The encounter of a singularity is accompanied by an extreme tilt limit 
(i.e., the fluid observer become asymptotically null with respect to the 
SH observer).
The singularity encountered by 
the fluid observer has the following properties:%
\footnote{Here, we refer to the properties of variables
associated with the orthonormal frame of the fluid observer.}  
\begin{itemize}
\item	The proper time needed to reach the singularity is finite,
\item	The Hubble scalar $H$ and some other kinematic quantities diverge,
\item	The matter density tends to zero (i.e., the singularity is not a 
	matter singularity).%
	\footnote{Since the source is a perfect fluid with a linear
	equation of state, this means that
	all components of the Ricci tensor also tend to zero.}
\end{itemize}
All components of the Weyl curvature tensor converge for some cases and 
at least one component diverges to infinity for other cases.
In this paper we shall further examine the behaviour of the components of 
the Weyl tensor (according to the fluid observer), with the aim to predict 
whether they converge or diverge.
In order to do so, we recall the following concept regarding the Weyl
curvature tensor.

\subsection*{The Weyl parameter}

The Weyl tensor is defined as the trace-free part of the Riemann 
curvature tensor and,
consequently, is not directly involved in the Einstein field equations.  
In some sense, the Weyl tensor describes the `free gravitational 
field' and therefore is of particular interest.
Penrose \cite{penrose} suggested that the Weyl tensor is related to a 
measure of a `gravitational entropy' and therefore could be used to shed 
light on the initial state of the Universe.  
In this context, the Weyl tensor and its Weyl scalars have been used to 
characterise different behaviours of non-tilted Bianchi models 
\cite{wa,gw,BHWeyl}.

Wainwright, Hancock and Uggla \cite{WHU99} studied the late-time 
dynamics
of Bianchi type VII$_0$ cosmologies with a non-tilted perfect fluid. 
They introduced a quantity $\mathcal{W}$ called the `Weyl parameter' (see
\cite{WHU99}, p 2580):
\be
\label{Weyl_param}
	\mathcal{W} = \frac{W}{H^2},\quad
	\text{where}
	\quad
	W^2 = \frac16(E_{ab}E^{ab} + H_{ab}H^{ab}).
\ee
They coined the term%
\footnote{In \cite{BHWeyl} the term `extreme Weyl 
dominance' was used for the case $\mathcal{W} \rightarrow \infty$ while 
`Weyl dominance' was used for the case when the Weyl curvature invariant 
dominated the Ricci invariant but $\mathcal{W}$ was bounded.}  `Weyl 
curvature dominance' to describe the phenomenon
in which $\mathcal{W} \rightarrow \infty$.
This means that the ratio of some components of the Weyl curvature tensor and
the square of the Hubble scalar tends to infinity.%
\footnote{It is worth noting that both $W$ and $H$ tend to zero in 
\cite{WHU99}.}
This phenomenon also occurs in Bianchi type VII$_0$ and VIII cosmologies
with a tilted perfect fluid \cite{VII,VIII}.

In \cite{WHU99,VII,VIII}, the Weyl parameter is defined with respect to
the SH observer.
However, the Weyl parameter can also be defined with respect to
different observers, and particularly the fluid observer in tilted Bianchi 
cosmologies.%
\footnote{The Weyl parameter, unlike the invariant Weyl scalars, is 
an observer-dependent quantity.}

Although the limits of the Weyl parameter along 
two congruences of worldlines are in principle different, 
we conjecture that in general they both tend to zero, constant, or 
infinity together.
We shall refer to this as `the Weyl parameters have the same 
convergence/divergence property'.
The argument is considered in two cases: whether or not the two 
congruences are 
asymptotically null with respect to each other ($v^2 \rightarrow 1$).

If the two congruences are not asymptotically null with respect to each
other ($v^2 \not\rightarrow 1$), then $\G = (1-v^2)^{-1/2}$ is bounded.
Then, from the boost formulae (\ref{boost_Eab})--(\ref{boost_Hab}) for the 
Weyl components, in general the Weyl components $\hat{C}_{abcd}$ and $C_{abcd}$ relative 
to two different congruences
have the same convergence/divergence property. Also recall from the boost
formula for the Hubble scalar $H$ (\cite{CHL06}, equation (B.15),
reproduced below)
\be
\label{boost_H}
        \hat{H} =       
        \frac13 \left[ \ehat_\mu(\Gv^\mu)
                        - \frac{\G}{\G+1} v^\mu \ehat_\mu(\G)
                        +3\G H  - 2 \G a_\mu v^\mu + \G \udot_\mu v^\mu
        \right]
\ee
that $\hat{H}$ and $H$ also have the same convergence/divergence
property. It then follows from (\ref{Weyl_param}) that generally the Weyl
parameters $\hat{\mathcal{W}}$ and $\mathcal{W}$ have the same
convergence/divergence property. 

In the more interesting case for this analysis, when the two congruences 
are asymptotically null with respect to each other ($v^2 \rightarrow 1$), 
then $\G 
\rightarrow \infty$
and, in general, $\hat{C}_{abcd}$ becomes of order $\G^2 C_{abcd}$. 
From (\ref{boost_H}) we see that $\hat{H}$ is of order $\G H$ as
$\G \rightarrow \infty$.
As a result, generally we have
\be
\label{CC}
	\frac{\hat{C}_{abcd}}{\hat{H}^2}
	\quad
	\text{is of order}
	\quad 
	\frac{C_{abcd}}{H^2};
\ee
i.e., the Weyl parameter $\hat{\mathcal{W}}$ of a second congruence is
generally of the same order as the Weyl parameter $\mathcal{W}$ of the
first congruence, and they therefore have the same convergence/divergence
property.
Exceptions are possible when cancellation occurs in the leading order
terms in all the Weyl components, as we will show occurs in the LRS 
Bianchi type V cosmologies.
Indeed, we will show that, when the argument works, we can use it to 
predict the type of singularity encountered by the fluid observer.

\section{Terminology}

We first review some terminologies about singularities and the asymptotic
dynamics of the Weyl tensor.

There are several classifications of singularities in the literature.
We are concerned with one particular classification regarding the
convergence/diver\-gence of the Weyl curvature tensor at the singularity.

\subsection*{Weyl singularity}

Collins and Ellis \cite[p 88]{CE79} used the terminology `Weyl
singularity' to describe the following: at least one component of the Weyl
curvature tensor (with respect to the orthonormal frame of the observer
who encounters the singularity) diverges.%
\footnote{Collins and Ellis actually used the terminology `conformal
singularity', but we feel `Weyl singularity' is more appropriate and
avoids confusion with the `conformal singularity' in the context of
isotropic singularities \cite{AT99,GCW92}.}

On the other hand, the matter density and all components of the Weyl 
tensor can converge as the singularity is approached,
while some kinematic quantities (most importantly, the Hubble scalar $H$) 
diverge. 
We shall call such a singularity a `kinematic singularity' when it is not 
a Weyl singularity or a matter singularity. 
In other words, relative to a congruence, a kinematic singularity is characterized 
by 
the blow-up of one or more kinematic variables in finite proper time,
while all components of the Weyl and Ricci tensors remain bounded.
This terminology is new.
A prototypical example of a kinematic singularity is the initial 
singularity encountered by the fundamental observers of the Milne universe 
\cite{Rindler}.

Later, in section~\ref{sec:discussion}, we shall discuss this classification
in the literature.

\subsection*{Weyl blow-up}

The concept of Weyl singularity combines two phenomena -- the blow-up
of Weyl and the occurrence of a singularity. To separate them, we
introduce the terminology `Weyl blow-up' to describe the phenomenon
when some components of the Weyl tensor diverge asymptotically with 
respect to a particular observer, regardless of the occurrence of 
singularity.%
\footnote{It is possible for Weyl blow-up to take an infinite
proper time to occur, in which case there is no singularity.}
Weyl blow-up can be simply stated as
\be
	W \rightarrow \infty,
\ee
where $W$ is defined in (\ref{Weyl_param}).
The phenomena $W \rightarrow const.$ or $W \rightarrow 0$ will be called 
`Weyl convergence'.
This terminology is new.
We have not encountered examples in tilted SH cosmologies in which $W$ 
is bounded but whose limit does not exist.

The Weyl scalars may or may not converge asymptotically when Weyl
blow-up occurs for one observer.
There are three scenarios.
In the trivial case where the Weyl tensor is identically zero, we
have Weyl convergence for all observers.
In the second case,
if at least one of the four Weyl scalars diverges, then we 
have Weyl blow-up for all observers.
In the third case, if all four Weyl scalars converge, then one observer 
may experience Weyl blow-up while another does not.
The third case is the case of interest here.

\subsection*{Future singularity and Weyl blow-up}

\begin{figure}[ht]
  \begin{center}
    \epsfig{file=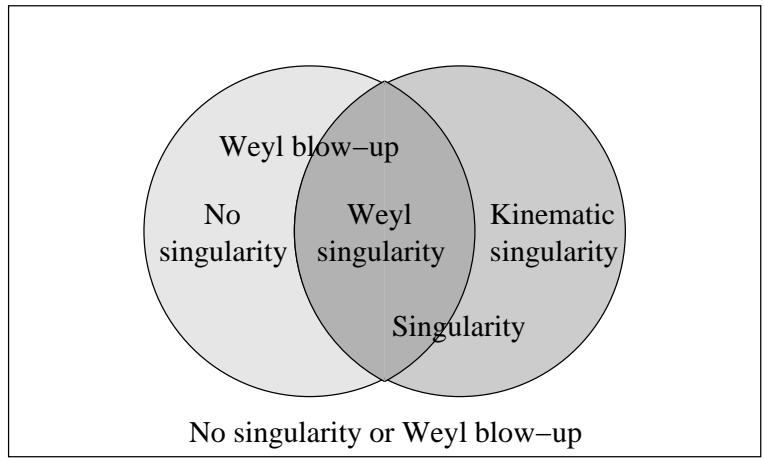}
    \caption{Four possible scenarios for the fluid observer at late 
times.}
    \label{fig:Venn}
\end{center}
\end{figure}

In the next section we shall examine tilted Bianchi cosmologies
that have an extreme-tilt sink. On approach to the sink, the
fluid observer may or may not encounter a future singularity, and may or
may not experience Weyl blow-up. 
We summarize the four possible scenarios with a Venn diagram (see 
figure~\ref{fig:Venn}).

We shall use the Weyl parameter to heuristically predict whether Weyl 
blow-up occurs if the
fluid observer encounters a future singularity; i.e., whether the
singularity is a Weyl singularity.
We will also give a special example in which the prediction breaks down.

\section{Typical Examples}

In this section we first examine three typical examples of tilted Bianchi 
cosmologies that have an extreme-tilt asymptotic regime at late times. For more details of the 
asymptotic dynamics of the tilt variable, the fluid cosmological time, and the fluid Hubble scalar, see 
\cite[Sections 2.1, 3.2.1, 3.2.2 and 5]{CHL06}.
The three examples illustrate different limits of the Weyl parameter of 
the fluid observer:
\[
	\hat{\mathcal{W}} \rightarrow 0,\quad
	\hat{\mathcal{W}} \rightarrow \infty,\quad
	\text{and}\quad
	\hat{\mathcal{W}} \rightarrow \text{const.,}
\]
respectively.

In all of the examples we have used the time parameter $\tau$ to be the 
dynamical time of the SH observer; i.e., $\tau$ is defined as 
\[
\frac{\mathrm{d}\tau}{\mathrm{d} t}=H,
\]
where $t$ and $H$ is the cosmological time and the Hubble parameter of 
the SH observer, respectively. We shall be using $\tau$ in the asymptotic 
expansions of both SH and fluid quantities. For those who would prefer to 
express the expansions in terms of the fluid cosmological time $t_{\rm 
fluid}$, simply use the relation (valid along each fluid worldline but not 
across the fluid congruence)
\[
	dt_{\rm fluid} = \frac{\sqrt{1-v^2}}{H} d\tau.
\]
For further details and discussion see \cite[page 3575]{CHL06}. 
In \cite{CHL06} we claimed that the limits of the kinematic variables 
along the SH congruence (or the normal congruence in the inhomogeneous 
example) and those along the fluid congruence are the same.
We take this opportunity to justify that claim.
The examples in \cite{CHL06} are special in that they have spatially 
homogeneous limits along the SH or normal congruence as $\tau$ tends to 
infinity. The fluid congruence is also special in that it must pass 
by all events passed by the SH or normal congruence. Thus, moving along a 
fluid worldline, the limit of a kinematic variable must be the same as the 
limit along an SH or normal worldline.

\subsection*{Irrotational Bianchi type V}

The dynamics of irrotational tilted Bianchi type V cosmologies was studied 
by Hewitt and Wainwright \cite{HW92}.
As the solutions approach the extreme-tilt Milne solution $M^-$
(which is a sink for $\gamma$ satisfying $\frac65 < \gamma < 2$)
at late times, the (orthogonal) SH observer sees
\be
\label{irro_1}
	H \approx H_0 e^{-\tau}	,\quad 
	W \approx H_0^2 |\Sigma_{-0}| e^{-4\tau} ,\quad
	\mathcal{W} \approx  |\Sigma_{-0}| e^{-2\tau},
\ee
as $\tau \rightarrow \infty$.%
\footnote{The asymptotic expansions for kinematic and matter variables are obtained by 
linearizing equations (2.10) and (2.11) in~\cite{HW92}, and solving them.
These are then substituted into equations (14) and (15) 
in~\cite[Appendix 3]{UEWE} to obtain the asymptotic expansions for the Weyl components.}
As a result
\be
\label{irro_2}
	H \rightarrow 0,\quad
	W \rightarrow 0,\quad
	\mathcal{W} \rightarrow 0,
\ee
for all $\frac65 < \gamma < 2$.

According to the (tilted) fluid observer, however,
the asymptotic expansions are (from the boost formulae (\ref{boost_H}) and 
(\ref{boost_Eab})--(\ref{boost_Hab}))
\begin{gather}
	\G \approx \G_0 \exp\left(
		\frac{5\gamma-6}{2-\gamma}\tau\right),\quad
	\hat{H} \approx \frac{4}{3(2-\gamma)} \G_0 H_0 
		\exp\left(\frac{2(3\gamma-4)}{2-\gamma}\tau\right),\quad
\\
	\hat{W} \approx 2\sqrt{2} \G_0^2 H_0^2 |\Sigma_{-0}|
		\exp\left(\frac{2(7\gamma-10)}{2-\gamma}\tau\right),\quad
	\hat{\mathcal{W}} \approx \frac{9\sqrt{2}}{8} (2-\gamma)^2
		|\Sigma_{-0}| e^{-2\tau},
\end{gather}
as $\tau \rightarrow \infty$, so
\be
\label{fluid_V}
	\begin{array}{ll}
	\hat{H} \rightarrow \infty & \text{for $\frac43 < \gamma <2$,}
\\
	\hat{W} \rightarrow \infty 
	& \text{for $\frac{10}{7} < \gamma <2$,}
\\
	\hat{\mathcal{W}} \rightarrow 0 
	& \text{for all $\frac65 < \gamma < 2$.}
	\end{array}
\ee
We notice that the convergence/divergence property of $H$ and $W$ is 
observer-dependent, while $\mathcal{W}$ tends to zero for both observers.
For the fluid observer, the $\gamma$ threshold for encountering a 
singularity is lower than that for Weyl blow-up. 
A singularity is not necessarily a Weyl singularity in this 
case (i.e., it could be a kinematic singularity).

\subsection*{Bianchi type VII$_0$}

The dynamics of tilted Bianchi type VII$_0$ cosmologies was studied
by Hervik \etal \cite{VII}.
As the solutions approach the extreme-tilt limit $\tilde{P}_4$
(which is a future attractor for $\gamma$ satisfying $\frac43 < \gamma < 2$) 
at late times, the SH observer sees
\be
	H \sim e^{-2\tau},\quad 
	W \sim e^{-3\tau},\quad
	\mathcal{W} \sim e^\tau,
\ee
as $\tau \rightarrow \infty$. As a result,
\be
        H \rightarrow 0,\quad
        W \rightarrow 0,\quad
        \mathcal{W} \rightarrow \infty,
\ee
for all $\frac43 < \gamma < 2$.

According to the fluid observer, however,
the asymptotic rates are
\begin{gather}
	\G \sim \exp\left(\frac{3\gamma-4}{2-\gamma}\tau\right),\quad
	\hat{H} \sim \exp\left(\frac{5\gamma-8}{2-\gamma}\tau\right),
\\
	\hat{W} \sim \exp\left(\frac{9\gamma-14}{2-\gamma}\tau\right),\quad
	\hat{\mathcal{W}} \sim e^\tau,
\end{gather}
as $\tau \rightarrow \infty$, so
\be
        \begin{array}{ll}
        \hat{H} \rightarrow \infty & \text{for $\frac85 < \gamma <2$,}
\\      
        \hat{W} \rightarrow \infty 
	& \text{for $\frac{14}{9} < \gamma <2$,}
\\
        \hat{\mathcal{W}} \rightarrow \infty 
	& \text{for all $\frac43 < \gamma < 2$.}
        \end{array}
\ee
Again, we notice that the convergence/divergence property of $H$ and $W$ 
is observer-dependent, while $\mathcal{W}$ tends to infinity (i.e., Weyl 
dominance) for both observers.
For the fluid observer, the $\gamma$ threshold for encountering a
singularity is higher than that for Weyl blow-up.
A singularity is necessarily a Weyl singularity in this case.%
\footnote{Tilted Bianchi type VIII cosmologies also exhibit the same 
qualitative behaviour (see \cite{VIII,CHLessay}).}

\subsection*{Bianchi type VII$_h$}

The dynamics of tilted Bianchi type VII$_h$ cosmologies was studied
in \cite{CH,CHV}.
As the solutions approach the extreme-tilt vacuum plane wave 
$\tilde{\mathcal{L}}_-(\text{VII} _h)$
(which is a sink for $\gamma$ satisfying $\frac6{5+2\Sigma_+} < 
\gamma < 2$, $ -\frac14< \Sigma_+ < 0$)
at late times, the SH observer sees
\be
	H \approx H_0 e^{(1-2\Sigma_+)\tau},\quad
	W \sim H^2,
\ee
as $\tau \rightarrow \infty$. As a result
\be
        H \rightarrow 0,\quad
        W \rightarrow 0,\quad  
        \mathcal{W} \rightarrow \text{const.},
\ee
for all $\frac6{5+2\Sigma_+} < \gamma < 2$.

According to the fluid observer, however,
the asymptotic rates are
\begin{gather}
	\G \approx G_0 \exp\left(
	\frac{((5+2\Sigma_+)\gamma-6)}{2-\gamma}\tau \right),
\\
	\hat{H} \sim \exp\left(
	\frac{2(3\gamma-2(2-\Sigma_+))}{2-\gamma}\tau \right),\quad
	\hat{W} \sim \hat{H}^2,
\end{gather}
as $\tau \rightarrow \infty$, so
\be
        \begin{array}{ll}
        \hat{H} \rightarrow \infty 
	& \text{for $\frac23 (2-\Sigma_+) < \gamma <2$,}
\\
        \hat{W} \rightarrow \infty 
	& \text{for $\frac23 (2-\Sigma_+) < \gamma <2$,}
\\
        \hat{\mathcal{W}} \rightarrow \text{const.}
	& \text{for all $\frac6{5+2\Sigma_+} < \gamma < 2$.}
        \end{array}
\ee
Again, we notice that the convergence/divergence property of $H$ and $W$
is observer-dependent, while $\mathcal{W}$ tends to a constant for both 
observers.
For the fluid observer, the $\gamma$ threshold for encountering a
singularity is equal to that for Weyl blow-up.
A singularity is necessarily a Weyl singularity in this case.

From the three typical examples above, we have formulated the following 
conjecture.

\paragraph{Conjecture:} In general (except in a zero-measure set of special solutions 
where the leading order term of each Weyl component vanishes),
 the convergence/divergence property of 
the Weyl parameter $\mathcal{W}$ is observer-independent. In the 
extreme-tilt limit, according to the fluid observer,
\begin{itemize}
\item	If $\hat{\mathcal{W}} \rightarrow 0$, then the $\gamma$ threshold 
	for 
	encountering a singularity is lower than that for Weyl 
	blow-up.
\item   If $\hat{\mathcal{W}} \rightarrow \infty$, then the $\gamma$ 
	threshold for   
        encountering a singularity is higher than that for Weyl
        blow-up.
\item   If $\hat{\mathcal{W}} \rightarrow$ const., then the $\gamma$ 
	threshold for
        encountering a singularity is equal to that for Weyl
        blow-up.
\end{itemize}

This conjecture works for typical examples, but may fail due to
simplification in special solutions. Let us present one such example.

\section*{Exceptional example: LRS Bianchi type V}

The LRS Bianchi type V cosmologies are a special case of the irrotational 
tilted Bianchi type V cosmologies. They were studied by Collins and 
Ellis~\cite{CE79}.

The shear component $\Sigma_-$ is zero in this class, and the SH observer 
does not see the asymptotic rates (\ref{irro_1}), but rather
\be
	H \approx H_0 e^{-\tau},\quad
	W \approx \frac52 H_0^2 |\Sigma_{+0}| e^{-6\tau},\quad
	\mathcal{W} \approx  \frac52 |\Sigma_{+0}| e^{-2\tau},
\ee
as $\tau \rightarrow \infty$. Nonetheless, (\ref{irro_2}) still holds:
\be
        H \rightarrow 0,\quad
        W \rightarrow 0,\quad
        \mathcal{W} \rightarrow 0,
\ee
for all $\frac65 < \gamma < 2$.

It turns out that there is only one independent 
component of the Weyl tensor $C_{abcd}$, namely
\be
	E_+ = (H+\sigma_+)\sigma_+ + \frac14 \frac{\gamma \mu}{G_+}V^2.
\ee
The boost formulae (\ref{boost_Eab})--(\ref{boost_Hab}) then imply that
$\hat{E}_+$ is the only nonzero component in any other frame, and that
\be
	\hat{E}_+ = E_+;
\ee
i.e., the leading order terms $\Gamma^2 C_{abcd}$ in $\hat{C}_{abcd}$ 
cancel. As a result, equation (\ref{CC}) fails, and this is a special 
example when general behaviour predicted by the conjecture fails.

Indeed, the fluid observer sees
\begin{gather}
        \G \approx \G_0 \exp\left(\frac{5\gamma-6}{2-\gamma}\tau
        \right),
\\
        \hat{H} \approx \frac{4}{3(2-\gamma)} \G_0 H_0
        \exp\left(\frac{2(3\gamma-4)}{2-\gamma}\tau\right),
\\ 
        \hat{W} = W \approx \frac52 H_0^2 |\Sigma_{+0}| e^{-6\tau},
\\
        \hat{\mathcal{W}} \approx \frac{45}{32}(2-\gamma)^2
        \G_0^{-2} |\Sigma_{+0}|
        \exp\left(\frac{-2(3\gamma-2)}{2-\gamma}\tau\right),
\end{gather}
as $\tau \rightarrow \infty$. As a result,
\be
        \begin{array}{ll}
        \hat{H} \rightarrow \infty & \text{for $\frac43 < \gamma <2$,}
\\
        \hat{W} \rightarrow 0 & \text{for all $\frac65 < \gamma <2$,} 
\\
        \hat{\mathcal{W}} \rightarrow 0 
	& \text{for all $\frac65 < \gamma < 2$.}
        \end{array}
\ee
Comparing with equation (\ref{fluid_V}), the quantity $\hat{W}$ never 
tends to infinity. 

\section{Discussion}\label{sec:discussion}

The future asymptotic dynamics in tilting SH Bianchi cosmologies
has recently been studied, with an emphasis on the physical properties 
of the models as 
experienced by the fluid observer \cite{CHL06,CHL06b}. In this paper
we have been primarily interested in the properties of the singularity 
that the
perfect fluid observer can encounter, with particular emphasis on the 
behaviour of the Weyl tensor (according to the fluid observer).
As we have seen in the examples, Weyl blow-up is not a good indicator
of a singularity. 
A better measure is the blow-up of kinematic variables.

We now discuss the work by
Collins and Ellis \cite{CE79} 
and
Ellis and King \cite{EK74} 
on Weyl blow-up at the singularity.
In Collins and Ellis \cite{CE79}, it is claimed that the future
singularity in the LRS Bianchi type V example is a Weyl singularity 
(i.e., a `conformal singularity', see page 97), 
in that some components of Weyl tensor diverge. 
This is not the case, for the only component of the Weyl tensor is $E_+$, 
and it tends to zero. 
All components of the Riemann tensor tend to zero at the singularity.
Some kinematic variables diverge, and are responsible for the singular 
behaviour.

Ellis and King \cite{EK74} classify singularities as
`curvature singularities', `locally extendible singularities' or
`intermediate singularities'.
Curvature singularities correspond to our Weyl singularities with the blow-up of at 
least one of the Weyl scalars; locally extendible singularities correspond to our 
kinematic singularities; and intermediate singularities correspond to our Weyl 
singularities with bounded Weyl scalars.
According to this classification, all of our examples of Weyl singularities are 
intermediate singularities.
We prefer the terminology `kinematic singularity' over 
`locally extendible singularity'
 since in our examples the fluid worldlines are 
certainly {\em not} extendible at the singularity, 
due to the blow up of the kinematic variables;
indeed, our examples are not `locally extendible' in the sense of 
Clarke \cite{C73}.

\subsection*{Conclusion}

In discussions regarding singularities, usually attention is focussed on the 
Weyl tensor components and the matter density. 
We have argued that in a cosmological context more emphasis should be put on the kinematic variables.

In this paper we have derived the boost formulae for the components of the 
Weyl tensor.
We have applied the boost formulae to establish that, for two observers 
that are asymptotically null with respect to each other, their respective 
Weyl parameter generally both tend to zero, constant, or infinity 
together.
We examined three classes of typical examples and one exceptional class.
Given the behaviour of the Weyl parameter, we can predict that the 
singularity encountered is a Weyl singularity or a kinematic singularity.
The examples suggest that the kinematic variables are more useful than 
the Weyl tensor components in indicating a singularity.

\section*{Acknowledgment}
This work was supported by an AARMS Postdoctoral Fellowship (SH)  
and the Natural Sciences and Engineering Research Council of
Canada (AC). 

\section*{Appendix: Boost formulae for the Weyl tensor components}

In \cite{CHL06} the boost formulae for the kinematic and the
matter variables are derived. Here we extend the formulae to include the
Weyl components.

Recall from \cite{CHL06} that the boost formulae for the orthonormal frame
vector fields are:
\begin{align}
        \ehat_0 &= \G \mathbf{e}_0 + \Gv^\mu \mathbf{e}_\mu
\\
        \ehat_\a &= \Gv_\a \mathbf{e}_0
                + B_\a{}^\mu \mathbf{e}_\mu
\end{align}
where
\be
        B_\a{}^\mu = \left[ \delta_\a{}^\mu
                + \frac{\G^2}{\G+1} v_\a v^\mu \right]
        \ ,\quad
        \G = \frac{1}{\sqrt{1-v^2}}
        \ ,\quad
        v^2 = v_\mu v^\mu
        \ .   
\ee
The Weyl curvature tensor $C_{abcd}$ can be decomposed into `electric' and
`magnetic' components as follows:
\be
        C_{\a0\b0} = E_{\a\b}\ , \quad
        C_{\a\b\gamma\delta} = - \epsilon^{\mu}{}_{\a\b}
                        \epsilon^{\nu}{}_{\gamma\delta} E_{\mu\nu}\ ,
\quad
        C_{\a\b\gamma0} = \epsilon^{\mu}{}_{\a\b} H_{\gamma\mu}\ .
\ee
Boosting the orthonormal frame results in the following boost formulae for
the Weyl components:
\begin{align}
\label{boost_Eab}
        \hat{E}_{\a\b} &= (2\G^2-1) E_{\a\b}
                - \frac{2\G^2(2\G+1)}{\G+1} v^\mu E_{\mu\la\a} v_{b\ra}
                + \frac{\G^4}{(\G+1)^2} E_{\mu\nu} v^\mu v^\nu
                        v_{\la\a} v_{\b\ra}
\notag\\
        &\quad
                + 2 \G^2 v_\mu \eps^{\mu\nu}{}_{(\a} H_{\b)\nu}
                - \frac{2\G^3}{\G+1}  v_\mu \eps^{\mu\nu}{}_{(\a} v_{\b)}
                        H_{\nu\gamma} v^\gamma
\\
\label{boost_Hab}
        \hat{H}_{\a\b} &= (2\G^2-1) H_{\a\b}
                - \frac{2\G^2(2\G+1)}{\G+1} v^\mu H_{\mu\la\a} v_{b\ra}   
                + \frac{\G^4}{(\G+1)^2} H_{\mu\nu} v^\mu v^\nu
                        v_{\la\a} v_{\b\ra}
\notag\\
        &\quad
                - 2 \G^2 v_\mu \eps^{\mu\nu}{}_{(\a} E_{\b)\nu}
                + \frac{2\G^3}{\G+1}  v_\mu \eps^{\mu\nu}{}_{(\a} v_{\b)}
                        E_{\nu\gamma} v^\gamma.
\end{align}

\end{document}